# Robust Diamond/β-Ga$_2$O$_3$ Hetero-p-n-junction Via Mechanically Integrating Their Building Blocks

Imteaz Rahaman[1], Hunter D. Ellis[1], and Kai Fu[1, a)]

[1]Department of Electrical and Computer Engineering, The University of Utah, Salt Lake City, UT 84112, USA

We report a novel approach for crafting robust diamond/β-Ga$_2$O$_3$ hetero-p-n-junctions through the mechanical integration of their bulk materials. This resulting heterojunction, with a turn-on voltage of ~2.7 V at room temperature, exhibits resilient electrical performance across a temperature spectrum up to 125°C, displaying minimal hysteresis—measuring as low as 0.2 V at room temperature and below 0.7 V at 125°C. Remarkably, the ideality factor achieves a record low value of 1.28, setting a new benchmark for diamond/ β-Ga$_2$O$_3$ heterojunctions. The rectification ratio reaches over $10^8$ at different temperatures. This effortlessly fabricated and remarkably resilient diamond/Ga$_2$O$_3$ hetero-p-n-junction pioneers a novel pathway for the exploration and fabrication of heterojunctions for ultra-wide bandgap semiconductors with substantial lattice mismatch and different thermal expansion coefficients.

---

[a)] Author to whom correspondence should be addressed. Electronic mail:  kai.fu@utah.edu

Wide bandgap (WBG) and ultra-wide bandgap (UWBG) semiconductors, such as SiC (3.2 eV),[1] GaN (3.39 eV),[2] β-Ga$_2$O$_3$ (4.5-4.8 eV),[3] and diamond (5.47 eV)[4], have undergone extensive exploration for applications in power electronics, extreme environment electronics, gas sensors, and UV detectors.[5,6] β-Ga$_2$O$_3$ stands out in high-power device applications, evident in Baliga's figure of merit comparison with Si (3000×), SiC (10×), and GaN (4×).[7] Moreover, β-Ga$_2$O$_3$ is gaining prominence due to its high breakdown field and suitability for mass production thanks to the availability of high-quality larger-size wafers.[8–10] Substantial research efforts have been dedicated to advancing Ga$_2$O$_3$ devices, leading to significant progress.[11–13]

Despite its potential, β-Ga$_2$O$_3$ faces challenges, notably the absence of p-type doping due to deep acceptor states and a flat valence band. Another hurdle is its exceptionally low thermal conductivity (10 to 30 W·m$^{-1}$·K$^{-1}$),[14,15] limiting the performance at high power and temperatures.[16] Addressing these challenges requires hetero-integration with other semiconductors possessing p-type characteristics and high thermal conductivity. Methods like wafer bonding, oxide deposition/sputtering, van der Waals bonding, and heteroepitaxy offer promising solutions. P-type diamond stands out as the best choice due to its outstanding thermal conductivity (2200 W·m$^{-1}$·K$^{-1}$), high critical breakdown field (10–20 MV·cm$^{-1}$), and well-established p-type characteristics.[17,18] Moreover, since the n-type doping of diamond and p-type doping of Ga$_2$O$_3$ pose challenges while the opposite doping types for both have been developed very well, fusing the p-type diamond and n-type Ga$_2$O$_3$ to create a diamond/Ga$_2$O$_3$ heterojunction can effectively overcome doping bottlenecks for both materials.

Kim *et al.* attempted to integrate mechanical exfoliated β-Ga$_2$O$_3$ flakes with diamond to create a solar-blind photodiode.[19] Matsumae *et al.* employed a low-temperature direct bonding method for β-Ga$_2$O$_3$ and single crystal diamond.[20] Another recent approach has grown polycrystalline β-

Ga$_2$O$_3$ on single crystal diamond.[21] Exfoliation poses a challenge due to device area limitations, and heteroepitaxial growth encounters obstacles like substantial lattice mismatch, substrate decomposition, material delamination, and intense residual stress in materials.[22,23] Ongoing research is actively addressing these challenges, seeking robust solutions for the construction of diamond/Ga$_2$O$_3$ heterojunctions.

In this study, we achieved the successful construction of a diamond/β-Ga$_2$O$_3$ hetero-p-n-junction through the mechanical integration of bulk p-type diamond and bulk n-type Ga$_2$O$_3$. The electrical performance of the hetero-p-n-junction remains robust up to 125 °C, with hysteresis lower than 0.7 V @ 1 µA. Notably, the ideality factor of the p-n junction is impressively low at 1.28, and the rectification ratio exceeds $10^8$. These findings highlight the significant potential of the mechanical integration approach, offering a promising avenue to simplify the fabrication process and facilitate the widespread application of diamond/Ga$_2$O$_3$ heterojunctions.

This heterojunction consists of a 450 µm thick p-type polycrystalline diamond doped with boron (B) at a concentration of ~ 4×10$^{20}$ cm$^{-3}$ and 650 µm thick n-type (001) β-Ga$_2$O$_3$ doped with tin (Sn) at a concentration of ~1×10$^{18}$ cm$^{-3}$. The materials were cleaned by a standard procedure, involving acetone, isopropanol, and deionized water. The hetero-p-n-junction was formed by aligning the diamond polished surface (< 30 nm) with the Ga$_2$O$_3$ polished surface (< 1 nm). Indium metals served as the metal contacts on the other surfaces for both samples followed by annealing at 300 °C for 30 s. The structure and schematic of the diamond/ β-Ga$_2$O$_3$ hetero-p-n-junction are illustrated in Fig. 1(a) and Fig. 1(b), respectively. The diamond/β-Ga$_2$O$_3$ hetero-p-n-junction benefits from a large conduction band discontinuity ($\Delta E_C$ = 3.5 eV), and valence band discontinuity ($\Delta E_V$ = 2.9 eV) by forming a type II heterojunction (Fig. 1(c)).[24] The testing setup for the diamond/β-Ga₂O₃ hetero-p-n-junction is depicted in Fig. 1(d), with the sample placed on a hot stage for evaluating its temperature-dependent performance.

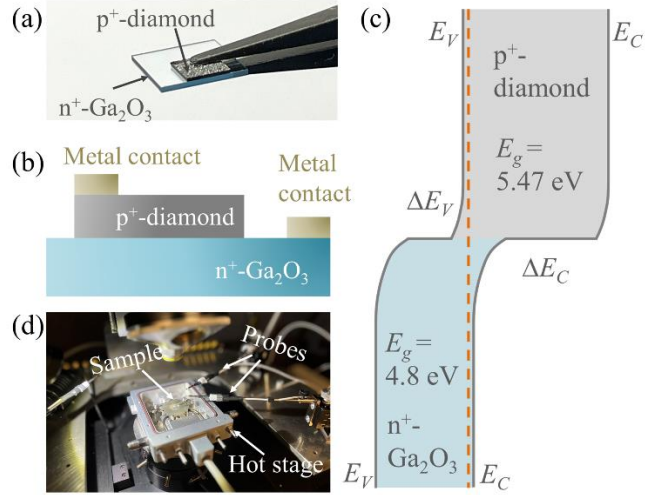

FIG. 1. Diamond/β-Ga₂O₃ hetero-p-n-junction structure. (a) Photo and (b) schematic of the diamond/Ga₂O₃ hetero-p-n-junction. (c) Energy band diagram illustrating the Type II hetero-p-n-junction. (d) Testing setup for evaluating its temperature-dependent performance.

The current-voltage ($I$-$V$) characteristics of the diamond/β-Ga₂O₃ hetero-p-n-junction were assessed under various temperatures. Multiple dual sweeps were conducted at each temperature to examine the repeatability and stability of the mechanically integrated heterojunction. The sweeping speed was consistent throughout, with a sweeping step of 100 mV, a source to measure delay of 5 ms, and measure window of 16.7 ms. As depicted in Fig. 2, the hysteresis observed in the $I$-$V$ curves over six dual sweeps is minimal, indicating excellent repeatability and stability of the hetero-p-n-junction at different temperatures when presented in linear scale. The forward current exhibits an increase with temperature, attributed to the enhanced diffusion constant and mobility (at high doping levels). It's noteworthy that the forward current surpasses that of reported diamond/Ga$_2$O$_3$ heterojunctions, primarily due to the easily achievable larger device area utilizing bulk materials.[11]

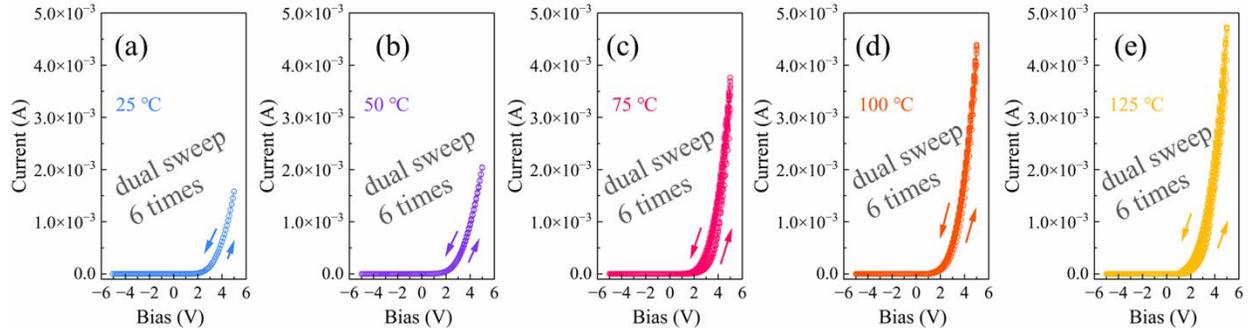

FIG. 2. Electrical performance at different temperatures. Current-voltage curves in linear scale for the diamond/β-Ga$_2$O$_3$ hetero-p-n-junction through six consecutive dual sweeps at (a) 25 °C, (b) 50 °C, (c) 75 °C, (d) 100 °C, and (e) 125 °C.

For a more in-depth analysis of hysteresis and performance in the heterojunction, semi-log plots are presented in Fig. 3. Notably, negligible hysteresis is observed at 25 °C and 50 °C, while a slight hysteresis becomes apparent above 75 °C. The *I-V* sweeps from forward bias to reverse bias exhibit higher current than their inverse counterparts, indicating a trapping process in the diamond, Ga$_2$O$_3$, or at the interface of diamond/Ga$_2$O$_3$. This trapping process weakens at low biases when trap levels are filled by carrier injection during the highest forward bias. The increased hysteresis at elevated temperatures aligns with our previous work, suggesting trapping and de-trapping processes at the interface.[6] Ideality factors of the heterojunction at different temperatures consistently show similar levels, while exact values are extracted from the lowest point in the linear region between 0 V and 1 V, as illustrated in Fig. 3.

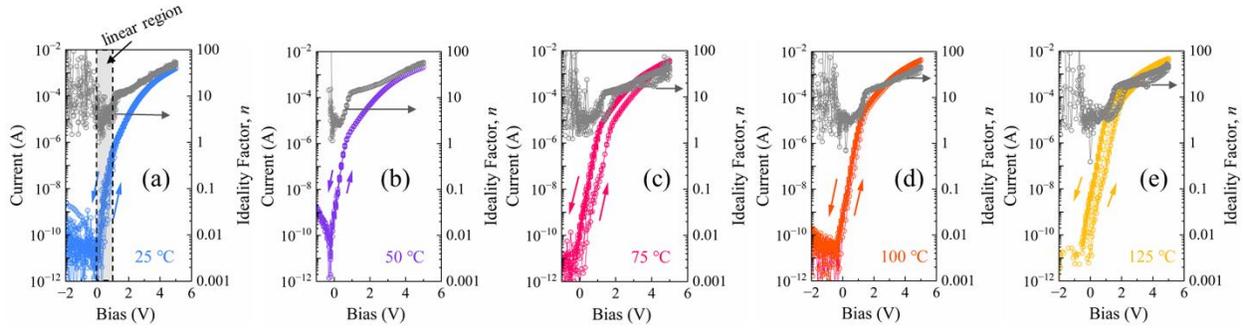

FIG. 3. Hysteresis and ideality factor. Current-voltage curves and ideality factors in semi-log scale for the diamond/β-Ga$_2$O$_3$ hetero-p-n-junction through six consecutive dual sweeps at (a) 25 °C, (b) 50 °C, (c) 75 °C, (d) 100 °C, and (e) 125 °C. The arrows indicate the dual sweep directions.

Figure 4 shows key parameters derived from the electrical performance of the heterojunction, encompassing forward current, reverse leakage current, rectification ratio, turn-on voltage, ideality factor, and hysteresis of $\Delta V$ at 1 µA. Notably, the diamond/Ga$_2$O$_3$ hetero-p-n-junction demonstrates a remarkable rectification ratio of $10^8$ at room temperature, further elevating at higher temperatures, except around 75 °C. The turn-on voltage, defined by two current levels (10 µA and 100 µA), ranges from 1 V to 3 V. As temperatures rise, the turn-on voltage decreases due to increased forward current. The ideality factor is impressively low at 1.28 at room temperature, representing a record low value for diamond/Ga$_2$O$_3$ heterojunctions. The hysteresis of $\Delta V$ at 1 µA varies between 0.1 V and 0.7 V for different temperatures, reaching as low as 0.2 V at room temperature. Table I presents a comparative analysis of ideality factor and rectification ratio for Ga$_2$O$_3$-based hetero-p-n-junctions. The proposed diamond/Ga$_2$O$_3$ hetero-p-n-junction exhibits outstanding overall performance. While breakdown voltage is not discussed here, it's worth noting that all bonded heterojunctions face a common challenge of low breakdown so far.

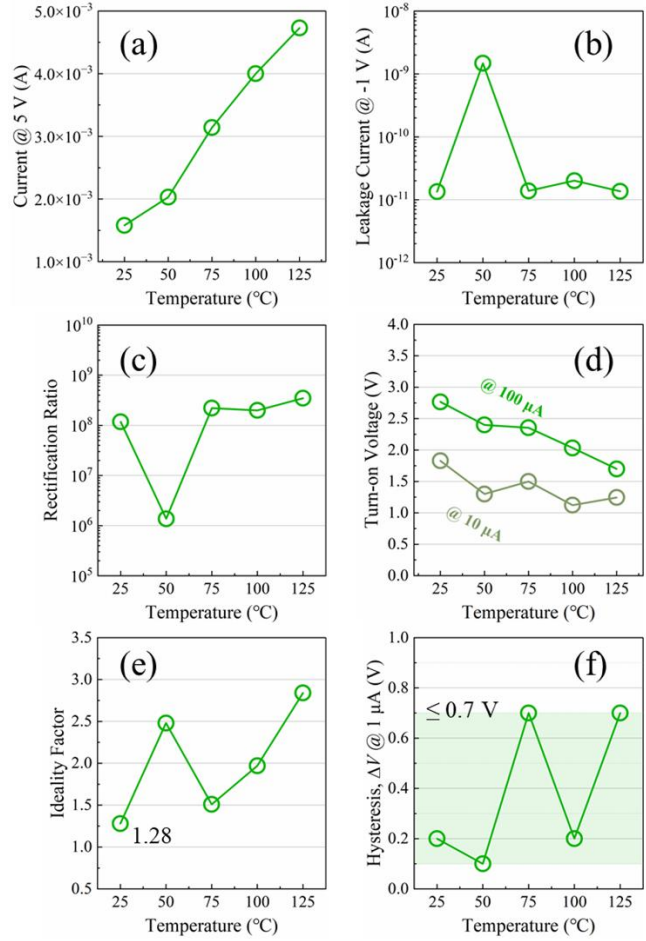

FIG. 4. Temperature dependence of the heterojunction parameters. (a) forward current (at 5 V), (b) leakage current (at −1 V), (c) rectification ration, (d) turn-on voltage determined by different current levels, (e) ideality factor, and (f) hysteresis (at 1 μA) at different temperatures.

Table I. Comparison of ideality factor and rectification ratio of β-Ga$_2$O$_3$ based heterojunctions.

| Heterojunction | Ideality factor | Rectification ratio |
|---|---|---|
| Oxide/Ga$_2$O$_3$ p-n | | |
| ZnCo$_2$O$_4$/ β-Ga$_2$O$_3$ [25] | ~2 | ~10$^9$ |
| NiO/ β-Ga$_2$O$_3$ [25] | ~2 | ~10$^9$ |
| Cu$_2$O/ β-Ga$_2$O$_3$ [26] | 1.31 | ~10$^7$ |
| SnO/ β-Ga$_2$O$_3$ [27] | 1.16 | 2×10$^8$ |
| α-Ir$_2$O$_3$/ α-Ga$_2$O$_3$ [28] | 4.41 | 3.12×10$^2$ |

| Semiconductor (other than diamond)/$Ga_2O_3$ p-n | | |
|---|---|---|
| SiC/ β-$Ga_2O_3$ [29] | >2.4 | >$10^6$ |
| GaAs/ β-$Ga_2O_3$ [30] | 1.23 | $8.04 \times 10^9$ |
| Si/ β-$Ga_2O_3$ [31] | 1.13 | $1.3 \times 10^7$ |
| GaN/ β-$Ga_2O_3$ [32] | 9.8 | ~$10^5$ |
| **Diamond/$Ga_2O_3$ p-n** | | |
| Diamond/ β-$Ga_2O_3$ [19] | 6.8 | $1.2 \times 10^7$ |
| Diamond/ β-$Ga_2O_3$ [24] | 2.7 | >$10^8$ |
| Diamond/ β-$Ga_2O_3$ [33] | 2.04 | ~$10^{10}$ |
| **This work** | **1.28** | **>$10^8$** |

This study unveils a novel approach for creating robust diamond/β-$Ga_2O_3$ hetero-p-n-junctions through the mechanical integration of their bulk materials. While typical ultra-wide bandgap (UWBG) heterojunctions encounter challenges due to significant lattice mismatch and pronounced differences in thermal expansion coefficients leading to fabrication difficulties and reliability concerns, the proposed easily fabricated and surprisingly resilient diamond/β-$Ga_2O_3$ hetero-p-n-junction, achieved through mechanical integration, provides a straightforward path for exploring and constructing UWBG heterojunctions. The heterojunction exhibited sensitivity to pressure but proved easily adjustable and optimizable during device packaging. An additional challenge lies in the effective contact area between bulk materials being smaller than the actual size in this work, resulting in low current density. This can be addressed by reducing the size of one of the materials. Ongoing efforts will focus on overcoming these challenges and elevating overall performance.

**AUTHOR DECLARATIONS**

**Conflict of Interest**

The authors have no conflicts to disclose.

**Author Contributions**

**DATA AVAILABILITY**

The data that support the findings of this study are available from the corresponding authors upon reasonable request.